\documentclass[12pt,preprint]{aastex}
\usepackage{lscape}

\def\spose#1{\hbox to 0pt{#1\hss}}

\def\farcs{\hbox{$.\!\!^{\prime\prime}$}}  
\def\lsim{\mathrel{\hbox{\rlap{\lower.55ex \hbox {$\sim$}}\kern-.0em\raise.4ex \hbox{$<$}}}} 
\def\gsim{\mathrel{\hbox{\rlap{\lower.55ex \hbox {$\sim$}}\kern-.0em\raise.4ex \hbox{$>$}}}}

\def\grb{GRB\,030329}

\shorttitle{Millimetric Observations of GRB 030329}
\shortauthors{Sheth et al.}

\begin{document}

\title{Millimeter Observations of GRB\, 030329: Continued Evidence for
  a Two-Component Jet}

\author{Kartik Sheth\altaffilmark{1,2}, Dale A. Frail
  \altaffilmark{3}, Stephen White\altaffilmark{4}, Mousumi
  Das\altaffilmark{4}, Frank Bertoldi \altaffilmark{5}, Fabian
  Walter\altaffilmark{3}, Shri R. Kulkarni\altaffilmark{1}, Edo
  Berger\altaffilmark{1}}

\altaffiltext{1}{Division of Mathematical \& Physical Sciences,
California Institute of Technology, Pasadena, CA 91125}

\altaffiltext{2}{Email: kartik@astro.caltech.edu}
\altaffiltext{3}{National Radio Astronomy Observatory, P.O. BOX 0,
  Socorro, NM 87801}
\altaffiltext{4}{Department of Astronomy, University of Maryland}
\altaffiltext{5}{Max-Planck-Institut f\"ur Radioastronomie, Auf dem
  H\"ugel 69, D--53121 Bonn, Germany}

\begin{abstract}
  We present the results of a dedicated campaign on the afterglow of
  \grb\ with the millimeter interferometers of the Owens Valley Radio
  Observatory (OVRO), the Berkeley-Illinois-Maryland Association
  (BIMA), and with the MAMBO-2 bolometer array on the IRAM 30-m
  telescope.  These observations allow us to trace the full evolution
  of the afterglow of \grb\ at frequencies of 100 GHz and 250 GHz for
  the first time.  The millimeter light curves exhibit two main
  features: a bright, constant flux density portion and a steep
  power-law decline. The absence of bright, short-lived millimeter
  emission is used to show that the GRB central engine was not
  actively injecting energy well after the burst. The millimeter data
  support a model, advocated by Berger et al., of a two-component
  jet-like outflow in which a narrow angle jet is responsible for the
  high energy emission and early optical afterglow, and a wide-angle
  jet carrying most of the energy is powering the radio and late
  optical afterglow emission.
\end{abstract}



\section{Introduction}\label{intro}

A link between long-duration gamma-ray bursts (GRBs) and the core
collapse of massive stars has long been claimed on observational (e.g.
Bloom, Kulkarni \& Djorgovski 2002\nocite{bkd02}; Bloom et
al.~2002\nocite{bkp+02}; Price et al.~2002\nocite{pbr+02}) and
theoretical grounds \citep{woo93,pac98b,mw99}, but the recent \grb\ 
has strengthened this association considerably.  Optical spectra taken
of this event \citep{smg+03,hsm+03,kdw+03} showed the usual power-law
continuum from the afterglow, superimposed upon which were lines
characteristic of a Type Ic supernova (SN).  Designated SN\,2003dh,
the brightness of this SNe and its broad line widths are strikingly
similar to another peculiar Ic SN\,1998bw \citep{pcd+01,mnn+02}, or
perhaps SN\,1997ef \citep{inn+00}.  Depending on the degree of
asphericity assumed in modeling the explosion, the derived kinetic
energies for SN\,1998bw and SN\,1997ef are in the range of 5-50 foe (1
foe=$10^{51}$ erg).  Such events are labeled ``hypernovae'', an
empirical classification to distinguish them from ordinary core
collapse SNe with energies of order 1 foe \citep{imn+98}.

While the non-relativistic component of the explosion (as traced by SN
\,2003dh) may have been hyper-energetic, the relativistic component
(as traced by \grb) seems to be sub-energetic. As noted by Granot,
Nakar \& Piran (2003)\nocite{gnp03} and Berger et
al.~(2003)\nocite{bkp+03}, if the sharp break in the optical light
curves at $t=0.55$ days \citep{pfk+03} is due to a jet-like outflow,
then the gamma-ray energy released E$_\gamma\sim$0.05 foe and the
X-ray luminosity (at t=10 hrs) L$_X\sim3\times 10^{43}$ erg s$^{-1}$.
These values of E$_\gamma$ and L$_X$ lie an order of magnitude or more
below which most GRBs are tightly clustered (Bloom, Frail \& Kulkarni
2003\nocite{bfk03}; Berger, Kulkarni \& Frail 2003\nocite{bkf03}).
Outliers in the energy/luminosity distribution are potentially
important for exploring the diversity of the GRB phenomena, and how
the central engine partitions the explosion energy.

Millimeter detections of gamma ray bursts (GRBs), while difficult to
achieve (e.g., Bremer et al. 1998\nocite{bkg+98}, Shepherd et al.
1998\nocite{sfkm98}), are a potentially powerful diagnostic of the
explosion energy. Since the peak of the synchrotron spectrum is
expected to pass through the millimeter band on a time scale of a day
or so after the burst (Sari, Piran \& Narayan 1998\nocite{spn98}),
such observations are useful for constraining the peak of the
spectrum, a quantity that is difficult to obtain by other means. When
combined with broadband data these millimeter observations have been
especially useful in deriving the kinetic energy of the outflow and
the density structure of the circumburst environment
\citep{gbb+00,bsf+00,hys+01,yfh+02}.

In this paper we present measurements of GRB\,030329 at 100 GHz made
with the millimeter interferometers of the Owens Valley Radio
Observatory (OVRO) and the Berkeley-Illinois-Maryland Association
(BIMA), and measurements at 250 GHz made with MAMBO-2 bolometer array
on the IRAM 30 m telescope.  At $z$=0.1686 \citep{cgh+03} this is the
closest known GRB and subsequently the flux density at millimeter
wavelengths was more than ten times brighter than any previous event.
These observations allow us to trace the full evolution of the
afterglow at millimeter wavelengths for the first time. In \S{2} we
discuss the observations and calibration issues, in
\S\ref{sec:results} and \S\ref{sec:discuss} we present the results in
the form of a light curve and discuss possible interpretations.

\section{Observations \& Data Reduction}

{\it Millimeter Interferometers:} We used the Owens Valley Radio
Observatory's millimeter array and the Berkeley-Illinois-Maryland
Association millimeter array to observe the \grb\ in a dedicated
campaign beginning 2003 March 30 through 2003 April 17.  An additional
observation was obtained on 2003 April 30.

For the first three nights, when the redshift of the GRB was unknown,
we tuned the local oscillator to the CS(2-1) line at a frequency of
97.98 GHz.  After that, we tuned to the CO(J=1-0) line redshifted to
z=0.1686.

We imaged the continuum flux using the analog correlator at OVRO which
consists of four 1 GHz bands around the LO frequency from $\pm$0.5-1.5
GHz and $\pm$2.5-3.5 GHz.  The array was in the H configuration and
the projected baselines ranged from 5.5 to 84.0 k$\lambda$ and the
single-sideband system temperatures ranged from 150 to 400 K.  The
complex instrumental gain was calibrated using the quasar 1159+292
every 15 minutes.  The flux of 1159+292 was determined using 3C\,273
since no planets were available for an absolute calibration; hence,
the resulting uncertainty in the overall flux scale is $\sim$15\%.  We
use a flux of 1.95 Jy for 1195+292 to calibrate all the data. The
calibrations were done with the MMA software package \citep{scc+93}
imaged using standard routines in MIRIAD (Sault, Teuben, \& Wright
1995)\nocite{stw95}.

At BIMA we used the 800 MHz digital correlator to map the GRB.  BIMA
was in the compact (C) configuration and projected baselines ranged
from 2.2--29.3 k$\lambda$.  The single-sideband temperatures ranged
from 200 to 400 K.  The same phase calibrator, 1159+292 was used to
calibrate the BIMA data. The results for the continuum flux
measurements are summarized in Table \ref{mmtab}.  At OVRO we used the
1 GHz digital spectrometer, and the new 8 GHz COBRA spectrometer, and
at BIMA we used the 800 MHz spectrometer both to search for a
redshifted CO spectral line in absorption; we did not detect it.

\begin{figure}
\epsscale{0.8}
\plotone{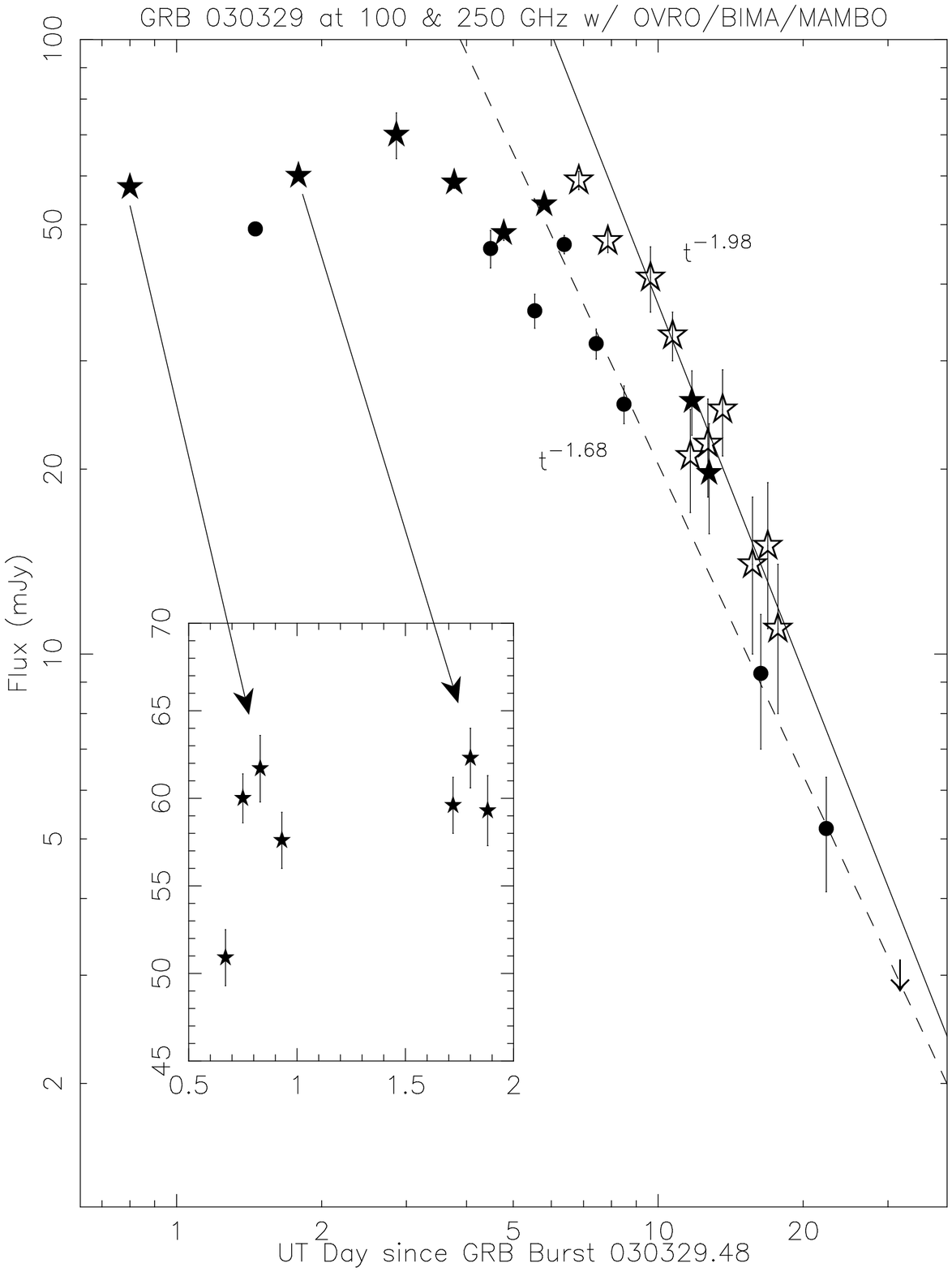}
\caption{Millimeter flux as a function of time on a logarithmic scale.
The stars indicate 100 GHz (3mm) data.  Filled stars represent OVRO
observations, and the open stars represent BIMA observations.  Filled circles 
represent the 250 GHz, IRAM 30-m MAMBO observations.  On two days
(2003 April 10 and April 11), both the BIMA and OVRO observatories
observed the GRB.  The measurements are consistent with each
other. The inset shows the measurements from the first two OVRO epochs
which have been subdivided into 2-hr intervals to look for flux
density variations.
\label{fig:mm}}
\end{figure}

Millimeter continuum flux from the GRB was detected at
$\alpha$(J2000)=10$^h$44$^m$49$^s$.95, $\delta$(J2000) =
21$^o$31$^\prime$ 17$\farcs$30.  This position is coincident with the
radio and optical afterglows.  The observing campaign is summarized in
Table \ref{mmtab}. On the first three days, we were also able to track
the evolution of the GRB flux within a scan because it was bright.
These data are listed in Table \ref{vartab}.  The light curve is
plotted in Figure \ref{fig:mm}.

{\it MAMBO 250 GHz observations:} The millimeter continuum
measurements were made using the 117-channel MAMBO-2 array (Kreysa et
al.\ 1999)\nocite{kgg+99} at the IRAM 30 m telescope on Pico Veleta
(Spain).  MAMBO-2 has a half-power spectral bandwidth of 210 and 290
GHz with an effective frequency of 250 GHz. The beam size on the sky
is 10.7 arcsec. The sources were observed with a single channel using
the standard on-off mode with the telescope secondary chopping in
azimuth by 32$^{\prime\prime}$ at a rate of 2$\,$Hz.

Observations of GRB\,030329 were done on eight different epochs
between March 30 and April 20, with integration times on sky ranging
between 5 and 17 minutes. The observing conditions were good, with the
exception of April 2, when the sky noise was unusually strong.  The
source was observed at elevations between 56 and 74 degrees, with line
of sight opacities at 250~GHz in the range 0.1 to 0.3.

The pointing and focus were checked frequently on the quasar 1043+241,
which is only 3 degrees away from the GRB. Usually the pointing was
found to be stable within $\sim 3^{\prime\prime}$. For the absolute
flux calibration a number of calibration sources were observed, mostly
CW-Leo, which is near the target. We used a flux calibration factor of
35,000 counts per Jansky, the one sigma uncertainty of which we
estimated to be $10\%$.

The data were analyzed using the MOPSI software package.  Correlated
noise was subtracted from each channel using the weighted average
signals from the surrounding channels. These data are listed in Table
\ref{mambotab} and the light curve is plotted in Figure \ref{fig:mm}.

\section{Results}\label{sec:results}

The 100 GHz and 250 GHz millimeter light curves in Figure \ref{fig:mm}
exhibit two main features: a bright, constant flux density portion and
a steep decline. A power-law fit to the decay gives $\alpha_R=-1.98$
(where $F_\nu\propto{t}^\alpha$) at 100 GHz and $\alpha_R=-1.68$ at
250 GHz. These values are in excellent agreement with that derived by
Price et al.~(2003)\nocite{pfk+03} for the optical decay beyond
$\sim$0.5 days ($\alpha_\circ=-1.97\pm 0.12$) and suggests a common
physical effect.

The second feature of the light curves in Figure \ref{fig:mm} is the
bright flux density plateau. In the first week after the burst we
derive a mean flux density at 100 GHz of 58 mJy. The emission is
constant during this time expect for a two day period starting April 2
where there is a small ($<20$\%) but significant drop in the flux
density followed by a rise to its mean level on April 5. Likewise, the
mean flux density at 250 GHz for the first 4 epochs prior to the decay
is 44 mJy. Small variations at 250 GHz during this time are likely due
to uncertainties in the calibration.


During the time that these millimeter measurements were made the the
optical light curve had a prominent bump at $\Delta{t}$=1.5 days with
the flux increasing by a factor of two.  There may be a second bump at
3.5 days, but this interpretation depends on whether we compare to a
power law with $\alpha=-2$ or $-1$.  In order to search for similar
variations at 100 GHz we subdivided each of our observing sessions
into 2-hour intervals. The results of this analysis are summarized in
Table \ref{vartab}. With the exception of a possible increase by about
10 mJy ($15-20\%$) at $t\approx 0.7$ d and $t\approx 2.8$ d, neither
of which corresponds to a change in the optical brightness, we find no
significant variations.

\section{Discussion}\label{sec:discuss}

Two explanations have been offered as to why \grb\ appears to be
under-energetic compared to other GRBs. Granot et
al.~(2003)\nocite{gnp03} have explained the observed steep decay in
the optical light curves of \grb\ at $t=0.55$ days \citep{pfk+03} in
terms of a standard model of a jet expanding into a constant density
medium (Sari, Piran \& Halpern 1999\nocite{sph99}). Note that such
steep temporal breaks $\Delta\alpha>1$ are not expected in spherical
outflows in general, nor are they expected in collimated outflows
expanding into a wind-blown medium (Sari et al.~1998; Kumar \&
Panaitescu 2000\nocite{kp00}). In their model the fluctuations in the
optical light curve (\S\ref{sec:results}) originate from ``refreshed
shocks'', in which slower moving ejecta catch up with the main shock
and re-energize it.  In this picture energy injection from the central
engine is not instantaneous (as commonly assumed) but episodic, with
most of the energy being carried by ejecta with the lower Lorentz
factors.

An unavoidable consequence of having the GRB central engine inject a
range of Lorentz factors $\Gamma$ over time, instead of a single high
Lorentz factor shell, is that each newly arrived shell at the shock
front modifies the afterglow spectrum, producing a short-lived reverse
shock, and shifts the combined spectrum to lower frequencies
(Rees \& M\'esz\'aros 1998; Kumar \& Piran 2000)\nocite{rees98,kp00c}.  According to the more detailed calculations of
Sari \& M\'esz\'aros (2000)\nocite{sm00}, which take into account
synchrotron self-absorption, there should be a significant enhancement
of the flux density at millimeter and submillimeter wavelengths.
Short-lived reverse shocks increase the peak flux density by $\Gamma$
and shift the peak to lower frequencies by a factor $\Gamma^{-2}$. For
the Lorentz factors given by Granot et al.~(2003) we would have
expected to see order of magnitude variations in the millimeter flux
decaying on timescales of a day or less. This key prediction is
contrary to what was seen (\S\ref{sec:results}) in our millimeter data
and thus our observations rule out the refreshed shock model.

Berger et al.~(2003)\nocite{bkp+03} proposed a two component jet model
based on the existence of two different jets breaks, one in the
optical light curves at 0.55 days and another best seen in the radio
light curves at 9.8 days. Single-jet fits to the radio data could not
explain the evolution of the optical emission before 1.5 days,
especially the sharp break at 0.55 days (Price et
al.~2003\nocite{pfk+03}). The millimeter observations presented here
are crucial in this respect since they define the peak of the
synchrotron spectrum and hence the overall normalization.  Lacking
such data, earlier claims of a two component outflow for GRB\,991216
\citep{fbg+00} are less secure. In this case Berger et al.~(2003)
define a narrow angle jet with $t_{NAJ}=0.55$ days and
$\theta_{NAJ}=0.09$ rad which is responsible for the early afterglow,
and a wide angle jet $t_{WAJ}=9.8$ days and $\theta_{WAJ}=0.3$ rad
which carries the bulk of the energy in the outflow and dominates the
optical and radio emission after $\sim 1.5$ days. It is likely that
the plateau, seen in the millimeter light curves (\S\ref{sec:results})
during the first week, is a combination of the falling flux density
(as t$^{-1/3}$) from the NAJ, and a rising flux density (as t$^{1/2}$)
from the WAJ (Sari et al.~1999\nocite{sph99}). 

Since in the two-component model the flux increase at 1.5 d is due to
the rise of the wide-jet component, which then evolves in the usual
fashion as $F_\nu\propto t^{-1}$, the subsequent possible fluctuations
discussed by Granot et al. (2003) and in \S\ref{sec:results} are
modest and no longer require refreshed shocks but instead could be
explained by variations in the circumburst density.

In summary, the millimeter observations presented here have been used
to distinguish between two equally compelling models for \grb\ and its
afterglow. The absence of bright, short-lived millimeter emission,
coincident with ``bumps'' in the optical light curve between 1 and 7
days after the burst, was used to show that the GRB central engine was
not actively injecting energy on this timescale. Instead, the
millimeter data support the proposed two-component jet. It is possible
that the true structure of GRB outflows are considerably more
complicated than the simple picture presented here (Zhang \&
M\'esz\'aros 2002\nocite{zm02}; Rossi, Lazzati \& Rees
2002\nocite{rlr02}; Perna, Sari \& Frail 2003\nocite{psf03}).  Models
of relativistic jets propagating out through the stellar progenitor
show that there may exist a large range in Lorentz factors in the
outflow (MacFadyen, Woosley \& Heger 2001\nocite{mwh01}; Zhang,
Woosley \& MacFadyen 2003\nocite{zwm03}), which decrease away from the
rotation axis as the degree of baryon entrainment increases.
Nonetheless, within the limits of the current data for \grb, there is
evidence for jet structure with at least two distinct components, with
the wider of the two carrying the bulk of the energy. This last point
is worth emphasizing since events like \grb/SN\,2003dh and
GRB\,980425/SN\,1998bw \citep{kfw+98} have shown that the group of
sub-energetic bursts may simply be an artifact of limited observations.
True calorimetry of GRBs must account for material at low Lorentz
factors \citep{bkp+03}, which typically are brightest at radio and
optical wavelengths.

\acknowledgements

GRB research at Caltech is supported by grants from NASA and the
National Science Foundation. The National Radio Astronomy Observatory
is a facility of the National Science Foundation operated under
cooperative agreement by Associated Universities, Inc. DAF thanks the
Astronomical Institute at the University of Amsterdam for their
hospitality during the time when this paper was written.
Research at the OVRO is partially funded by NSF grant AST-9981546, and
at BIMA by NSF grant AST-9981289.  


\clearpage

\clearpage

\begin{deluxetable}{rllrrcl}
\tabletypesize{\scriptsize}
\tablecaption{Observations \label{mmtab}}

\tablehead{\colhead{UTDay} &\colhead{Beginning UT} &\colhead{End UT} &\colhead{Flux} &\colhead{1$\sigma$ Rms} &\colhead{Observatory} &\colhead{Comments}}
\startdata
03Mar30 & 02:32 &  11:10 & 57.5 & 0.8 & OVRO & \\
03Mar31 & 03:48 & 09:16 &  60.0 & 0.9 & OVRO & \\
03Apr01 & 05:49 & 10:47 &  70   & 6   & OVRO & (uv-avg, terrible weather) \\
03Apr02 & 01:22 & 10:39 &  58.5 & 1.3 & OVRO & \\
03Apr03 & 01:59 & 10:38 &  48.4 & 1.3 & OVRO & \\
03Apr04 & 04:38 & 09:01 &  53.9 & 0.9 & OVRO & \\
03Apr05 & 04:43 & 10:30 &  59 & 2 & BIMA & \\
03Apr06 & 04:49 & 11:33 &  47 & 2 & BIMA & \\
03Apr07 & ... & ... & ... & ... & OVRO & No data, Bad weather \\
03Apr08 & 02:56 & 03:17 & 41 & 5 & BIMA & \\
03Apr09 & 04:27 & 04:49 & 33 & 3 & BIMA & \\
03Apr10 & 03:02 & 03:29 & 21 & 4 & BIMA & \\
        & 04:55 & 06:10 & 25.8 & 3.1 & OVRO & \\
03Apr11 & 03:44 & 04:08 & 22 & 4 & BIMA & \\
        & 05:02 & 06:17 & 19.7 & 4.0 & OVRO & \\
03Apr12 & 01:56 & 02:25 & 25 & 4 & BIMA & \\
03Apr13 & ...   & ...   & ...& ... & BIMA & No data, Bad weather \\
03Apr14 & 04:07 & 04:30 & 14 & 4 & BIMA & \\
03Apr15 & 08:30 & 09:18 & 15 & 4 & BIMA & \\
03Apr16 & 04:40 & 05:28 & 11 & 3 & BIMA & \\
03Apr17 & 03:48 & 04:28 & ... & ... & BIMA & No data, Bad weather \\
 ...    & ...   & ...   & ... & ... & ...  & Apr 18-30, no obs \\
03Apr30 & 04:06 & 08:15 & ND  & 3   & BIMA & Upper limit 3 mJy/bm \\
\enddata
\end{deluxetable}

\begin{deluxetable}{lrllrrcl}
\tabletypesize{\scriptsize}
\tablecaption{Early Time Evolution\label{vartab}}

\tablehead{\colhead{Day} &\colhead{Beginning UT} &\colhead{End UT} &\colhead{Flux} &\colhead{1$\sigma$ Rms} &\colhead{Observatory} &\colhead{Comments}}
\startdata
03Mar30 & 02:32 &  04:32 & 50.9 & 1.6 & OVRO & \\
03Mar30 & 04:32 &  06:32 & 60.0 & 1.4 & OVRO & \\
03Mar30 & 06:32 &  08:32 & 61.7 & 1.9 & OVRO & \\
03Mar30 & 08:32 &  11:10 & 57.6 & 1.6 & OVRO & \\
\\
03Mar31 & 03:48 & 05:48 & 59.6 & 1.6 & OVRO & \\
03Mar31 & 05:48 & 07:48 & 62.3 & 1.7 & OVRO & \\
03Mar31 & 07:48 & 09:16 & 59.3 & 2.0 & OVRO & \\
\\
03Apr02 & 01:22 & 04:22 & 55.4 & 1.8 & OVRO & \\
03Apr02 & 04:22 & 07:22 & 66.5 & 2.1 & OVRO & \\
03Apr02 & 07:22 & 10:39 & 54.6 & 2.4 & OVRO & \\
\enddata
\end{deluxetable}

\begin{deluxetable}{lcccccc}
\tabletypesize{\scriptsize}
\tablecaption{250 GHz MAMBO Observations\label{mambotab}}
\tablehead{
\colhead{Day} &
\colhead{Beginning UT} &
\colhead{Int (sec)} &
\colhead{Flux} &
\colhead{1$\sigma$ Rms} &
\colhead{Comments}
}
\startdata
03Mar30  & 22:21 & 1031 &  49.2 &  1.1 &  elev.=74$^\circ$, $\tau$=0.21\\
03Apr02  & 06:55 &  156 &  45.7 &  3.2 &  elev.=62$^\circ$, $\tau$=0.36\\
03Apr02  & 23:07 &  308 &  36.2 &  2.3 &  elev.=71$^\circ$, $\tau$=0.29\\
03Apr04  & 00:29 &  310 &  41.6 &  1.6 &  elev.=56$^\circ$, $\tau$=0.13\\
03Apr04  & 20:39 &  312 &  46.4 &  1.6 &  elev.=65$^\circ$, $\tau$=0.13\\
03Apr05  & 21:54 &  306 &  32.0 &  1.8 &  elev.=74$^\circ$, $\tau$=0.10\\
03Apr06  & 23:17 &  310 &  25.5 &  1.8 &  elev.=67$^\circ$, $\tau$=0.10\\
03Apr14  & 19:30 &  468 &   9.3 &  2.3 &  elev.=60$^\circ$, $\tau$=0.26\\
03Apr20  & 19:15 &  970 &   5.2 &  1.1 &  elev.=62$^\circ$, $\tau$=0.24\\
\enddata
\end{deluxetable}
\end{document}